\documentclass[prd,showpacs,11pt,nofootinbib,aps, floatfix]{revtex4-1}

\usepackage{amsmath}
\usepackage{amssymb}
\usepackage[dvips]{graphicx}
\usepackage{hyperref}
\usepackage{url}

\begin{document}

\title{Observational constraints on Rastall's cosmology}

\author{C.~E.~M. Batista}
\email{cedumagalhaes@hotmail.com}
\affiliation{Physics Department, Universidade Estadual de Feira de Santana, Brazil}
\author{J.~C. Fabris}
\email{fabris@pq.cnpq.br}
\author{O.~F. Piattella}
\email{oliver.piattella@pq.cnpq.br}
\author{A.~M. Velasquez-Toribio}
\email{alan.toribio@ufes.br}
\affiliation{Physics Department, Universidade Federal do Esp\'irito Santo, Brazil}


\begin{abstract}
Rastall's theory is a modification of General Relativity, based on the non-conservation of the stress-energy tensor. The latter is encoded in a parameter $\gamma$ such that $\gamma = 1$ restores the usual $\nabla_\nu T^{\mu\nu} = 0$ law. We test Rastall's theory in cosmology, on a flat Robertson-Walker metric, investigating a two-fluid model and using the type Ia supernovae Constitution dataset. One of the fluids is pressureless and obeys the usual conservation law, whereas the other is described by an equation of state $p_x = w_x\rho_x$, with $w_x$ constant. The Bayesian analysis of the Constitution set does not strictly constrain the parameter $\gamma$ and prefers values of $w_x$ close to $-1$. We then address the evolution of small perturbations and show that they are dramatically unstable if $w_x \neq -1$ and $\gamma \neq 1$, i.e. General Relativity is the favored configuration. The only alternative is $w_x = -1$, for which the dynamics becomes independent from $\gamma$.
\end{abstract}

\maketitle


\section{Introduction}

The nature of dark matter (DM) and dark energy (DE) is one of the most important open questions today in physics. There are strong observational evidences indicating that about $95\%$ of the universe is under the form of DM ($\approx 25\%$) and DE ($\approx 70\%$), but no direct detection has been reported until now. The usual candidates to DM (e.g. neutralinos and axions) and DE (e.g. cosmological constant, quintessence) lead to very robust scenarios, but at same time they must face theoretical and observational issues. For recent reviews on the subject, see for example \cite{Li:2011sd, Caldwell:2009ix, Bertone:2004pz}.
\par
Among the alternatives to the standard description of the dark sector there is the possibility of a modification of gravity theory on large scales, see \cite{Clifton:2011jh} for a recent review on the subject. An example are the so-called $f(R)$ theories, which are based on the inclusion of non-linear curvature terms in the Einstein-Hilbert action \cite{DeFelice:2010aj}. Other possibilities are the Unified Dark Matter models, where DM and DE are treated as a single entity \cite{Kamenshchik:2001cp, Gorini:2007ta, Piattella:2009da, Piattella:2009kt, Bertacca:2010mt, Campos:2012ez}, models in which DM is treated as a viscous component \cite{Zimdahl:1996ka, Colistete:2007xi, HipolitoRicaldi:2010mf, Piattella:2011bs} and models of interaction in the dark sector (i.e. exchange of energy between DM and DE) \cite{Zimdahl:2001ar, Zimdahl:2005bk}.
\par
Another example, the one we pursue in this paper, is to touch one of the cornerstone of the gravity theory: the usual conservation laws for matter components. This kind of formulation was introduced by Rastall some 40 years ago \cite{Rastall:1973nw, Rastall:1976uh}, and has been recently investigated in a cosmological context, giving some interesting results concerning the dynamics of the dark sector \cite{Fabris:2011rm, Batista:2011nu, Fabris:2011wz, Daouda:2012ig, Fabris:2012hw}.
\par
Rastall's motivation for modifying the usual conservation laws is based on the fact that the latter have been directly tested only locally or in a weak-field regime. On the other hand, the introduction of covariant derivatives imply, in some sense, an exchange of energy between matter and the gravitational field. Hence, in general, non-trivial generalizations of the conservation law are in principle possible. Besides, particle production in curved spacetimes is a central issue of quantum field theory on such spaces \cite{Birrell:1982ix}. Therefore, one may regard modifications of the usual (classical) conservation laws as effective, semi-classical approaches to such phenomenon.
\par
Rastall's proposal is the following:
\begin{equation}
{T^{\mu\nu}}_{;\mu} = \kappa R^{;\nu}\;,
\end{equation}
where the semicolon denotes the covariant derivative and $\kappa$ is a (dimension-full) free parameter.
The above relation can be rewritten as
\begin{equation}\label{Rastmod}
{T^{\mu\nu}}_{;\mu} = \frac{\gamma - 1}{2} T^{;\nu}\;,
\end{equation}
where $T$ is the trace of the stress-energy tensor and $\gamma$ is now a dimensionless free parameter. When $\kappa = 0$, then $\gamma = 1$ and the usual conservation law (and thus General Relativity) is recovered.
\par
In a one-fluid model, it is possible to redefine the energy-momentum tensor in order to recover the usual conservation law \cite{Fabris:1998hr}. In this sense, Rastall's theory is just a redefinition of the fluid equation of state. However, in a multi-fluid case the modification introduced by Rastall opens possibilities of non-trivial interactions among the different components. In \cite{Batista:2011nu}, this feature is used to investigate the consequences of the modified conservation law for a model of the dark sector of the universe: the authors investigate a two-fluid model, one of them being pressureless matter $p_{m} = 0$, whereas the other obeying the vacuum energy equation of state $p_{x} = - \rho_{x}$. Assuming that the matter component obeys the usual conservation law, then the vacuum energy conservation law is affected by the presence of matter, via Eq.~\eqref{Rastmod}. The main result of \cite{Batista:2011nu} indicates that the model is completely equivalent to the $\Lambda$CDM at the background
and linear perturbations levels. There is just one striking difference: DE may now agglomerate. This fact could have important consequences at the non-linear level, which is a regime where the $\Lambda$CDM faces some difficulties \cite{Perivolaropoulos:2008ud}.
\par
In the model studied in \cite{Batista:2011nu}, the equivalence with the $\Lambda$CDM at background and linear perturbations levels implies that no constraints on the parameter $\gamma$ can be established using the corresponding observational tests. Such constraints are, on the other hand, in principle possible using non-linear data. In the present paper our goal is to verify to what extent the model studied in reference \cite{Batista:2011nu} is a favorable configuration. To do this, we repeat the analysis made there but with the $x$ component now being described by a more general equation state, i.e. $p_{x} = w_x\rho_{x}$, with $w_x$ constant.
\par
We show that in this case the parameter $\gamma$ appears explicitly in the background and linear perturbation equations. Therefore, using type Ia supernovae data, we calculate its probability distribution function (PDF) along with the one for the matter density parameter $\Omega_{m0}$, and for the equation of state parameter $w_x$. The analysis show that supernovae date do not constrain strictly $\gamma$ whereas $w_x \sim -1$ is favored. Then, considering the evolution of small perturbations we find a dramatic instability if $w_x \neq -1$ and $\gamma \neq 1$. This is a result which favors General Relativity, i.e. $\gamma = 1$. On the other hand, another possibility seems to be viable, i.e. $w_x = -1$, for which the dynamical equations turn independent of $\gamma$ and which is the case investigated in \cite{Batista:2011nu}, which thus seems to be favored.
\par
The paper is organized as follows. In Sec.~\ref{Sec:RastThe} we present Rastall's theory, deriving some cosmological considerations. In Sec.~\ref{Sec:Obsconstr} we confront the predictions of Rastall's theory with data from the type Ia supernovae of the Constitution set. In Sec.~\ref{Sec:Pertan} we tackle the question of the evolution of small perturbations and, finally, in Sec.~\ref{Sec:ResDisc} we present our conclusions.


\section{Rastall's theory}\label{Sec:RastThe}

According to Eq.~\eqref{Rastmod}, Einstein equations must be modified as follows:
\begin{equation}
G_{\mu\nu} = R_{\mu\nu}- \frac{1}{2}g_{\mu\nu}R= 8\pi G\left( T_{\mu\nu}- \frac{\gamma-1}{2} g_{\mu\nu}T\right)\;,
\end{equation}
in order to be compatible with Bianchi identities (we use $c = 1$ units hereafter).

We consider a two-fluid model. The first component mimics baryons and DM, i.e. has negligible effective pressure, whereas the second describes an exotic dark component responsible for the acceleration of the universe, and has an equation of state of the form $p_x = w_x\rho_x$, with $w_x$ constant.
We assume the matter component to conserve as usual. This is important in order to have matter agglomeration required to form the local structures. The field equations become
\begin{eqnarray}
\label{R}
R_{\mu\nu} &=& 8\pi G\left[T^{x}_{\mu\nu}+ T^{m}_{\mu\nu} - \frac{1}{2}(2-\gamma)g_{\mu\nu}(T^{x}+T^{m})\right]\;,\\
\label{n} T^{\mu\nu}_{x}{}_{;\mu} &=& \left(\frac{\gamma-1}{2}\right)\left(T^{;\nu}_{x} +  T^{;\nu}_{m}\right)\;, \qquad
T^{\mu\nu}_{m}{}_{;\mu} = 0\;,
\end{eqnarray}
where subscripts or superscripts $x$ and $m$ denote the DE and the matter component, respectively. On large scales we assume the universe to be described by the spatially flat Friedmann-Lemaitre-Robertson-Walker (FLRW) line element,
\begin{equation}\label{FLRWmet}
ds^{2}= dt^{2}- a(t)^{2}\left[dr^{2} + r^{2}(d\theta^{2}+\sin^{2}\theta d\varphi^{2})\right]\;,
\end{equation}
where $a(t)$ is the scale factor. The perfect fluid energy-momentum tensor has the following form:
\begin{equation}
 T^{\mu\nu}_{x,m}=(\rho_{x,m} + p_{x,m})u^{\mu}u^{\nu} - p_{x,m}g^{\mu\nu}\;.
\end{equation}
Then, the field equations take on the following form:
\begin{eqnarray}
\left(\frac{\dot{a}}{a}\right)^{2} &=& \frac{4\pi G}{3}\left\{\rho_{x}\left[3-\gamma - 3(1-\gamma)w_{x}\right] + (3-\gamma)\rho_{m}\right\}\;,
\label{1420}\\
\frac{\ddot{a}}{a} &=& \frac{4\pi G}{3}\left\{\left[3(\gamma-2)w_{x} -\gamma\right]\rho_{x}  - \gamma\rho_{m}\right\}\;,\\
\dot{\rho}_{x} + 3\frac{\dot{a}}{a}(1+ w_{x})\rho_{x} &=& \frac{\gamma-1}{2}\left[(1-3w_{x})\dot{\rho}_{x}  + \dot{\rho}_{m}\right]\;,
\label{1419}\\
\label{rhomeq}\dot{\rho}_{m} + 3\frac{\dot{a}}{a}\rho_{m} &=& 0\;,
\end{eqnarray}
where the dot denotes derivative with respect to the cosmic time $t$. Note how the expansion history of the universe depends of the parameters $\gamma$ and $w_x$. However, it is not difficult to show that, if $w_x = -1$, then neither $H^2$ nor $\ddot{a}/a$ depend on $\gamma$. This is the case investigated in \cite{Batista:2011nu}, where the model has a background expansion identical to the $\Lambda$CDM one.

The evolution of $\rho_m$ and $\rho_x$ as functions of the scale factor are easily determined by solving Eqs.~\eqref{1419} and \eqref{rhomeq}, which yields to:
\begin{eqnarray}
\rho_x &=& \rho_{de0} a^{-\frac{6(1+w_x)}{2-(\gamma-1)(1-3w_{x})}} + \frac{(1-\gamma)\rho_{m0}}{2w_x + (\gamma-1)(1-3w_x)}a^{-3}\;,
\label{3a}\\
\rho_{m} &=& \frac{\rho_{m0}}{a^{3}}\;.
\end{eqnarray}
From Eq.~\eqref{3a} one can infer a constraint on $\gamma$ by requiring $\rho_x$ to be positive. To be conservative, we require the second term, which acts as a contribution to matter, to be positive and then one can find that, assuming $w_x < 0$,
\begin{equation}\label{gammaconstr}
 1 \le \gamma < \frac{1 - 5w_x}{1 - 3w_x}\;.
\end{equation}
The right limit is not included because it makes the denominator vanishing. For $w_x = -1$, , i.e. the case investigated in \cite{Batista:2011nu}, one obtains $1 \le \gamma < 3/2$. The first term in Eq.~\eqref{3a} has an exponent which, in principle, could also be positive, mimicking thus a phantom component. If we want to avoid this to occur, we can find more constraints. We still assume $w_x$ to be negative, but now we have to distinguish between the two cases $w_x < - 1$ and $-1 < w_x < 0$ (the case $w_x = -1$ ``kills'' the exponent). Thus, the exponent of the first term in Eq.~\eqref{3a} is negative for:
\begin{eqnarray}
 \gamma < \frac{3(1 - w_x)}{1 - 3w_x}\;, \qquad -1 < w_x < 0\;,\\
 \gamma > \frac{3(1 - w_x)}{1 - 3w_x}\;, \qquad w_x < -1\;.
\end{eqnarray}
The first constraint is contained in Eq.~\eqref{gammaconstr}, which is more restrictive. The second one combined with Eq.~\eqref{gammaconstr} gives:
\begin{equation}\label{gammaconstrnophantom}
 \frac{3(1 - w_x)}{1 - 3w_x} \le \gamma < \frac{1 - 5w_x}{1 - 3w_x}\;.
\end{equation}
Thus, we come to an interesting result: in Rastall theory it is possible to have DE with an equation of state $w_x < -1$, without necessarily it being phantom. The parameter $\gamma$ has to be chosen according the above constraint and the value $\gamma = 1$ is evidently not included. It is also important to stress that, when
\begin{equation}
 \gamma < \frac{1 - 5w_x}{1 - 3w_x}\;,
\end{equation}
then the exponent is larger than $-3$ and therefore it is sub-dominant in the past with respect to the matter contribution.
\par
Note that the exotic fluid density can be split into two fluids, one of which behaves as matter. Moreover, we have denoted as $\rho_{m0}$ the present value (i.e. for $a = 1$) of the matter density, and as $\rho_{de0}$ the present value of that part of the $x$-fluid which behaves as DE, being the total density of the $x$-fluid today
\begin{equation}
\rho_{x0} = \rho_{de0} + \frac{(1-\gamma)\rho_{m0}}{2w_x + (\gamma-1)(1-3w_x)}\;.
\end{equation}
The present-time density parameters of the DE and of matter are given by
\begin{equation}
\Omega_{de0}= \frac{8\pi G\rho_{de0}}{3H^{2}_{0}}\;, \quad \Omega_{m0}= \frac{8\pi G\rho_{m0}}{3H^{2}_{0}}\;,
\label{baa}
\end{equation}
and in a flat universe are related as follows, by using Friedmann equation~\eqref{1420}:
\begin{equation}
\Omega_{de0} = \frac{2}{3 -\gamma - 3(1 - \gamma)w_x} - \Omega_{m0}\left[\frac{(1 - \gamma)}{2w_x + (\gamma - 1)(1 - 3w_x)} + \frac{3 - \gamma}{3-\gamma-3(1-\gamma)w_x}\right]\;.
\label{36}
\end{equation}
In the case $\gamma = 1$, the above relation recovers the corresponding General Relativity one: $\Omega_{de0} + \Omega_{m0} = 1$.


\section{Observational Constraints with type Ia supernovae}\label{Sec:Obsconstr}

We consider type Ia supernova data (the Constitution set \cite{Hicken:2009dk}, which uses the SALT filter) and perform a Bayesian analysis. In more detail, we calculate first the $\chi^2$ function, defined as follows:
\begin{equation}
 \chi^2 = \sum_{i = 1}^N\frac{\left[x_i^{th}(P) - x_i^{ob}\right]
 ^2}{\sigma_i^2}\;,
\end{equation}
where $N$ is the total number of observational data $x_i^{ob}$ with uncertainty $\sigma_i$ and $x_i^{th}(P)$ are their theoretical predictions, which depend on a set $P = \left\{p_1,p_2,..\right\}$ of parameters. Assuming the data to be independent Gaussian random variables, see for example \cite{Riess:1998cb}, and flat priors on the parameters, the posterior probability function is constructed from the $\chi^2$ function as follows
\begin{equation}
 \mbox{PDF}(P) = e^{-\chi^2(P)/2}\;,
\end{equation}
For our model, we have four free parameters: $h$, $\Omega_{m0}$, $\gamma$ and $w_x$. However, $h$ can be handled in a special way and it can be excluded, leaving just three free parameters.

Type Ia supernovae data consist in the distance modulus $\mu$, i.e.
\begin{equation}
\mu \equiv m_{obs}(z_{i}) - M = 5 \log_{10}\left(\frac{d_{L}}{\rm Mpc}\right) + 25\;,
\end{equation}
where $m_{obs}$ is the apparent magnitude, $M$ is the absolute magnitude and $d_{L}$ is the luminosity distance
\begin{equation}
 d_L(z) = c(1 + z)\int_0^{z}\frac{dz'}{H(z')}\;.
\end{equation}
The distance modulus can also be written as
\begin{equation}
\mu = 5\log_{10}D_{L}(z) + \mu_{0}\;,
\end{equation}
where $D_{L}= (H_{0}d_{L})/c$ is the Hubble-free luminosity distance and $\mu_{0}$ is the zero point offset, defined by
\begin{equation}
\mu_{0}=5\log_{10}\left(\frac{cH_{0}^{-1}}{\mbox{Mpc}}\right) +25 = 42.39 - 5\log_{10} h\;.
\end{equation}
For the Rastall's model considered here, the Hubble parameter is given by
\begin{eqnarray}
H(z)= \biggr\{\Omega_x(z)\biggr[\frac{3 - \gamma - 3(1 - \gamma)w_x}{2}\biggl] + \frac{(3 - \gamma)}{2}\Omega_m(z)\biggl\}^{1/2},
\end{eqnarray}
with
\begin{eqnarray}
\Omega_x(z) &=& \Omega_{de0}\,(1 + z)^\frac{6(1 + w_x)}{2 - (\gamma - 1)(1 - 3w_x)} + \frac{(1 - \gamma)}{2w_x + (\gamma - 1)(1 - 3w_x)}\Omega_m(z),\\
\Omega_m(z) &=& \Omega_{m0}(1 + z)^3,
\end{eqnarray}
$\Omega_{de0}$ and $\Omega_{m0}$ satisfying the condition (\ref{36}).

We employ data from the so-called Constitution set \cite{Hicken:2009dk}, which includes 397 distance moduli, out of which 100 come from the new low-z CfA3 sample and the rest from the Union set. Both samples have a redshift range of $0.015 \leq z \leq 1.55$. The main improvement of the Constitution sample is the inclusion of a larger number of nearby $(z < 0.2)$ supernovae, thus reducing the statistical uncertainty.

\begin{figure}[ht]
\begin{center}
\includegraphics[width=0.4\columnwidth]{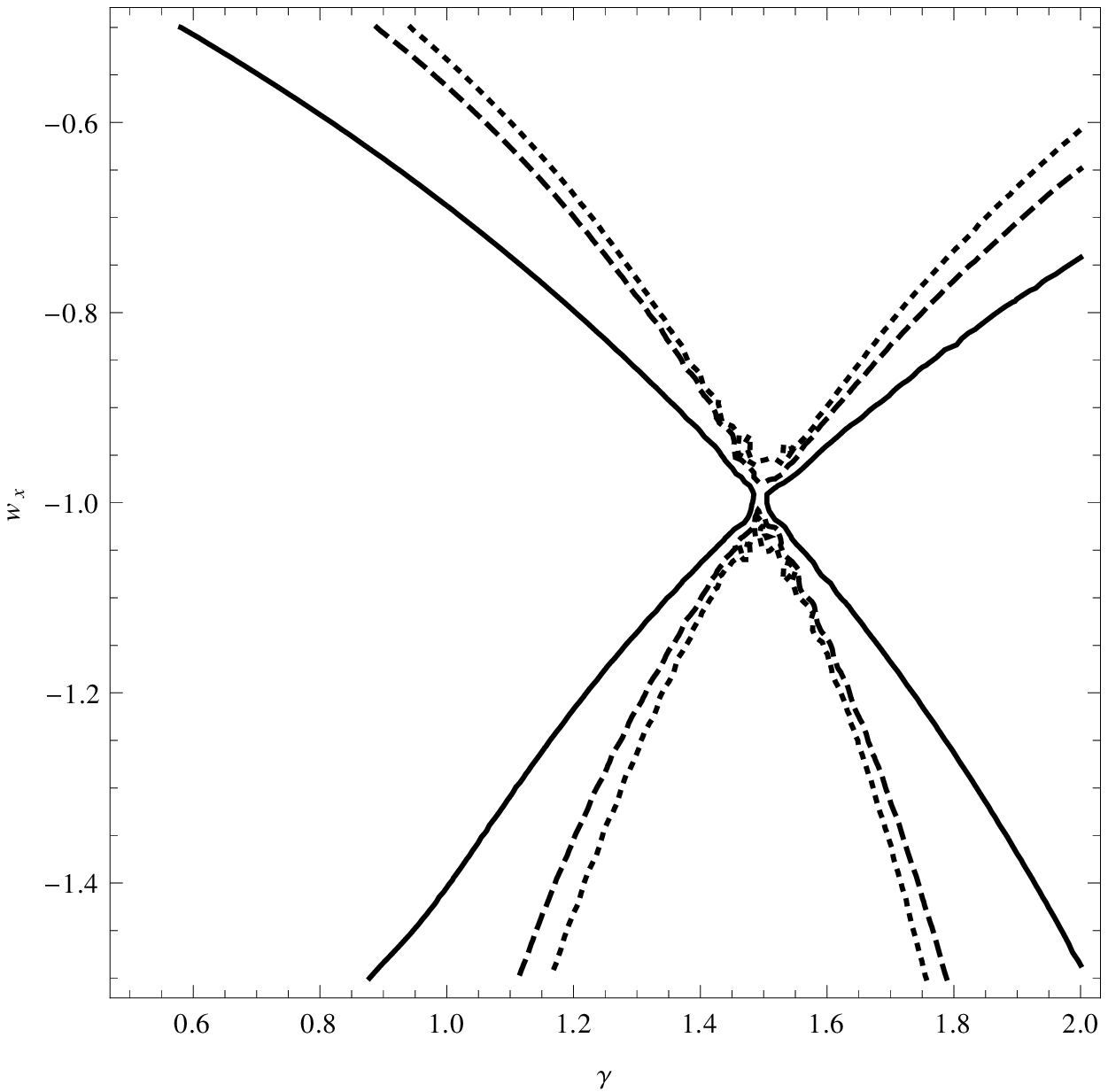}\includegraphics[width=0.4\columnwidth]{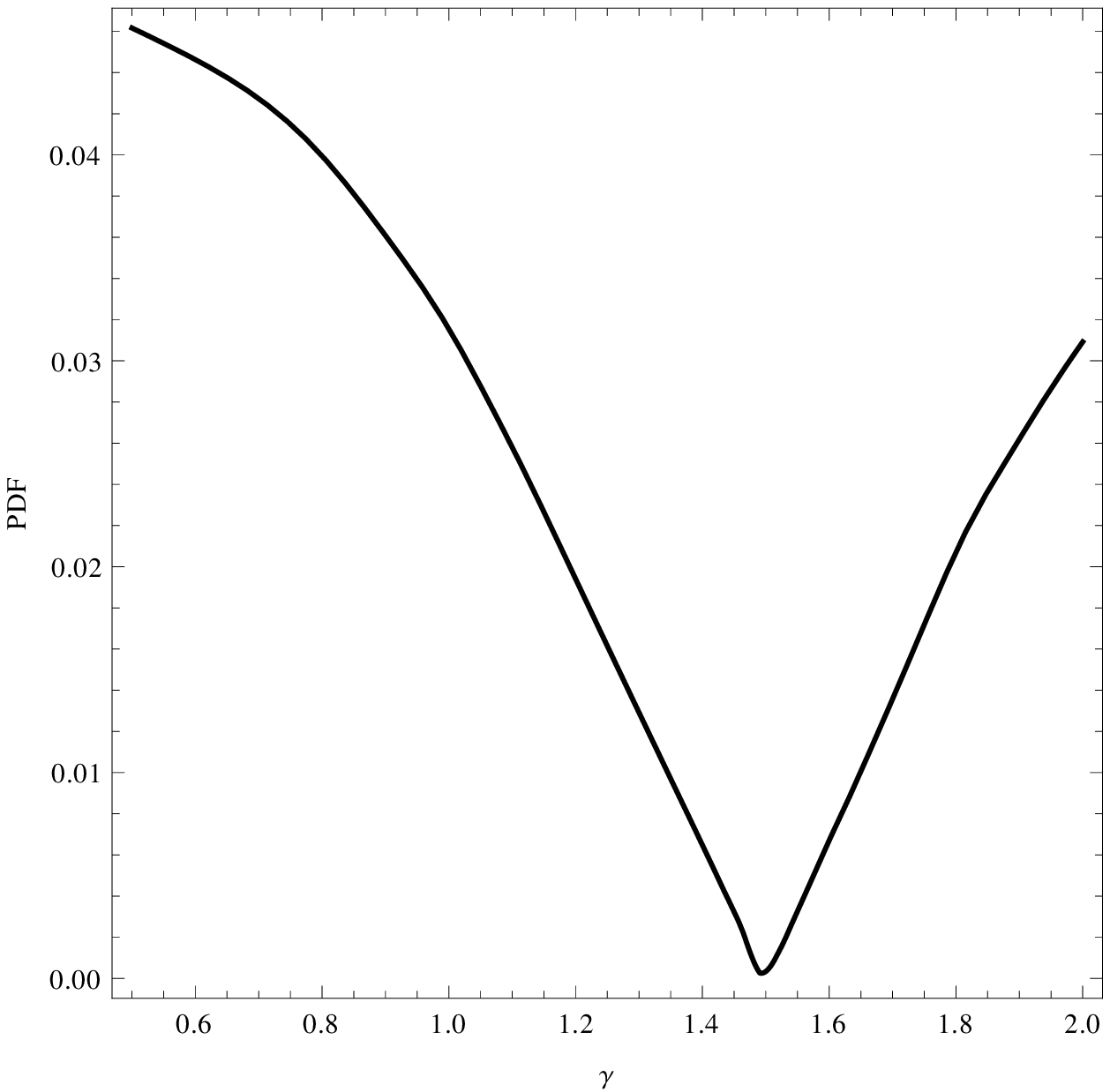}\\
\includegraphics[width=0.4\columnwidth]{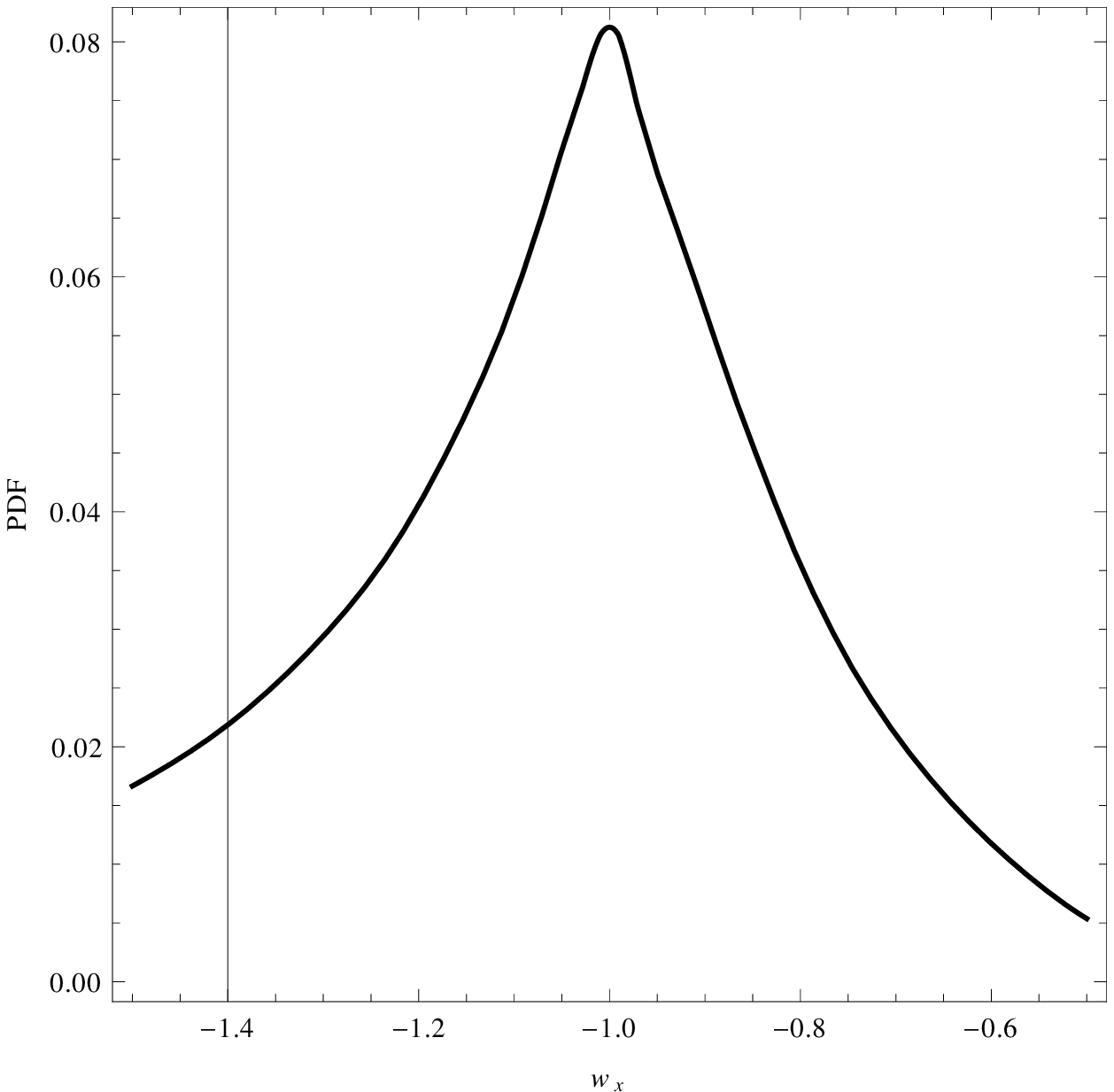}\includegraphics[width=0.4\columnwidth]{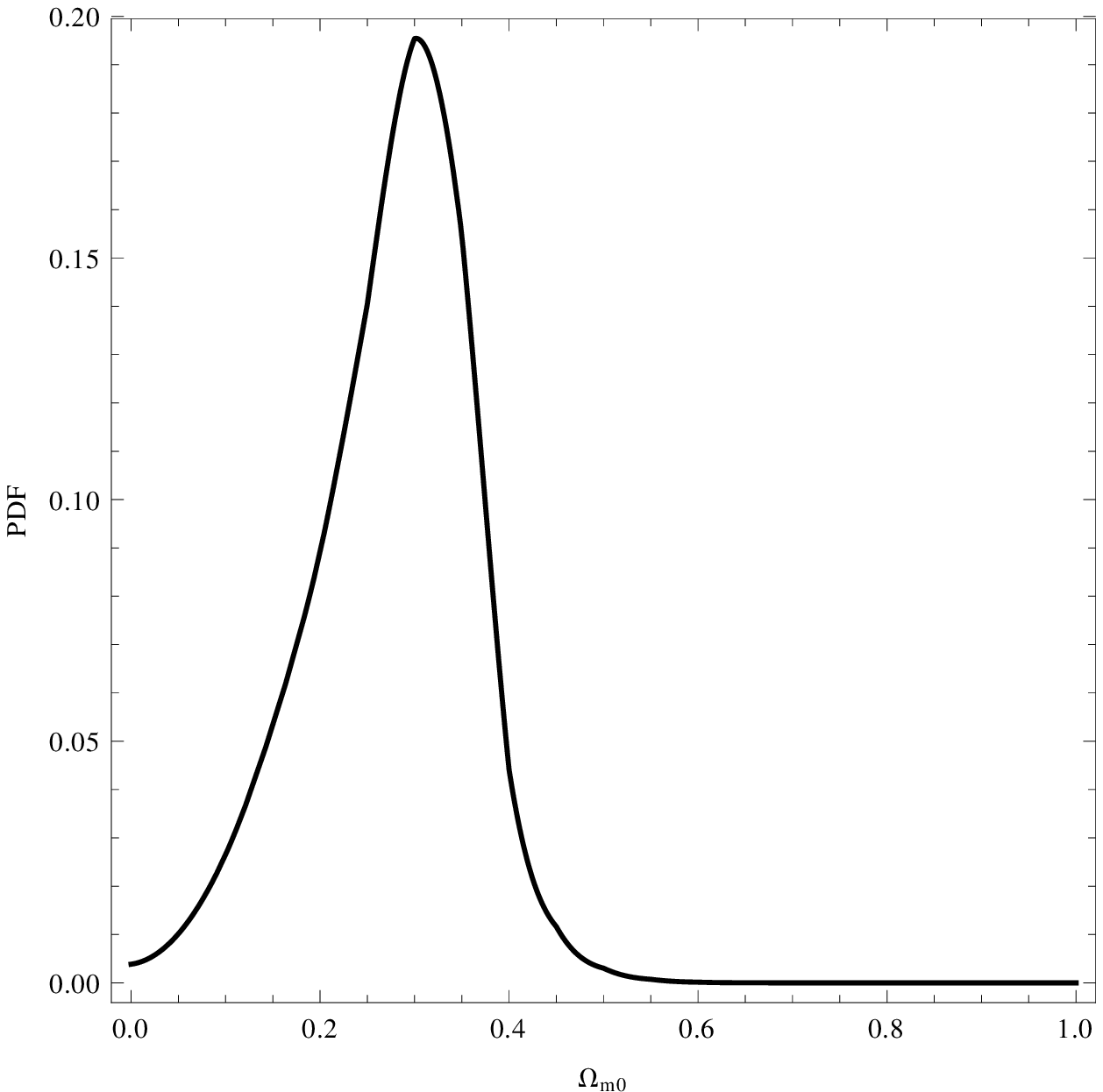}
\end{center}
\caption{Upper left panel: contour plots for the 68\%, 95\% and 99\% CL (solid, dashed, dot-dashed lines, respectively) in the plane $\gamma - w_x$. Upper right panel: PDF for $\gamma$. Lower left panel: PDF for $w_x$. Lower right panel: PDF for $\Omega_{m0}$.}
\label{fig1}
\end{figure}

The chi-squared we employ for the type Ia supernovae test is then
\begin{equation}\label{chi2sn}
\chi^{2}_{\rm SNIa}(P) = \sum_{i=1}^{397}{\frac{\left[\mu_{th}(P,z_{i}) - \mu_{ob,i}(z_{i})\right]^{2}}{\sigma_{ob,i}^{2}}}\;,
\end{equation}
where $P =\left\{\Omega_{m0},w_x,\gamma,\mu_0\right\}$. The chi-square can be already minimized with respect to $\mu_{0}$, since the latter appears as a linear dependence. Expanding Eq.~\eqref{chi2sn} with respect to $\mu_{0}$, we obtain
\begin{equation}
\chi^{2}(p)_{\rm SNIa} = A(p) + 2\mu_{0}B(p)+ \mu_{0}^{2}C(p)\;,
\end{equation}
where $p =\left\{\Omega_{m0},w_x,\gamma\right\}$ and with the following definitions:
\begin{eqnarray}
A(p) &\equiv&\sum_{i}^{n}{\frac{\left[\mu_{th}(p) - \mu_{obs}\right]}{\sigma_{i}^2}^{2}}\;,\\
B(p) &\equiv& \sum_{i}^{n}{\frac{\mu_{th}(p) - \mu_{obs}}{\sigma^2_{i}}}\;,\\
C(p) &\equiv& \frac{1}{\sigma_{i}^{2}}\;.
\end{eqnarray}
Therefore, the chi-square is minimum for $\mu_{0} = -B(p)/C(p)$, giving thus
\begin{equation}
 \bar{\chi}^{2}_{\rm SNIa}(p) = A(p) - \frac{B^{2}(p)}{C(p)}\;.
\end{equation}
As a side remark, for the single-fluid model Rastall's theory has been confronted against supernova data (Union sample) leading to results that can be competitive to the $\Lambda$CDM data under special conditions \cite{Capone:2009xm}. In \figurename{~\ref{fig1}} we show the contour plots for $\gamma$ and $w_x$ and the single PDF for the three parameters.

Note that in the upper left panel of \figurename{~\ref{fig1}} the credible region extends to very small and to very large values of $\gamma$. As the PDF in the right upper panel suggests, values about $\gamma = 3/2$ are less probable. The latter number is the right hand side of Eq.~\eqref{gammaconstr} for $w_x = -1$, which is the favored value (see the left lower panel of \figurename{~\ref{fig1}}). When $\gamma = 3/2$ and $w_x = -1$ the right hand side of Eq.~\eqref{3a} diverges (this is also the reason for the numerical issues appearing about that point in the right upper panel of \figurename{~\ref{fig1}}).
\par
The mininum value for the $\chi^2$ implies $\gamma = 1.41336$, $w_x = -1.00337$ and $\Omega_{m0} = 0.296399$. When the PDF is marginalized, the maximum PDF is reached at
$w_x = -0.999546$ and $\Omega_{m0} = 0.31133$, while for $\gamma$ grows at the extreme of the interval after a minimum around $\gamma = 3/2$.
Just for a comparison, for the Union 2 dataset, the $\chi^2$ minimum implies $\gamma = 1.51993$, $w_x = -0.937786$ and $\Omega_{m0} = 0.123751$, values near those of the Constitution
dataset, except for $\Omega_{m0}$.


\section{Perturbative analysis}\label{Sec:Pertan}

\begin{figure}[ht]
\begin{center}
\includegraphics[width=0.4\columnwidth]{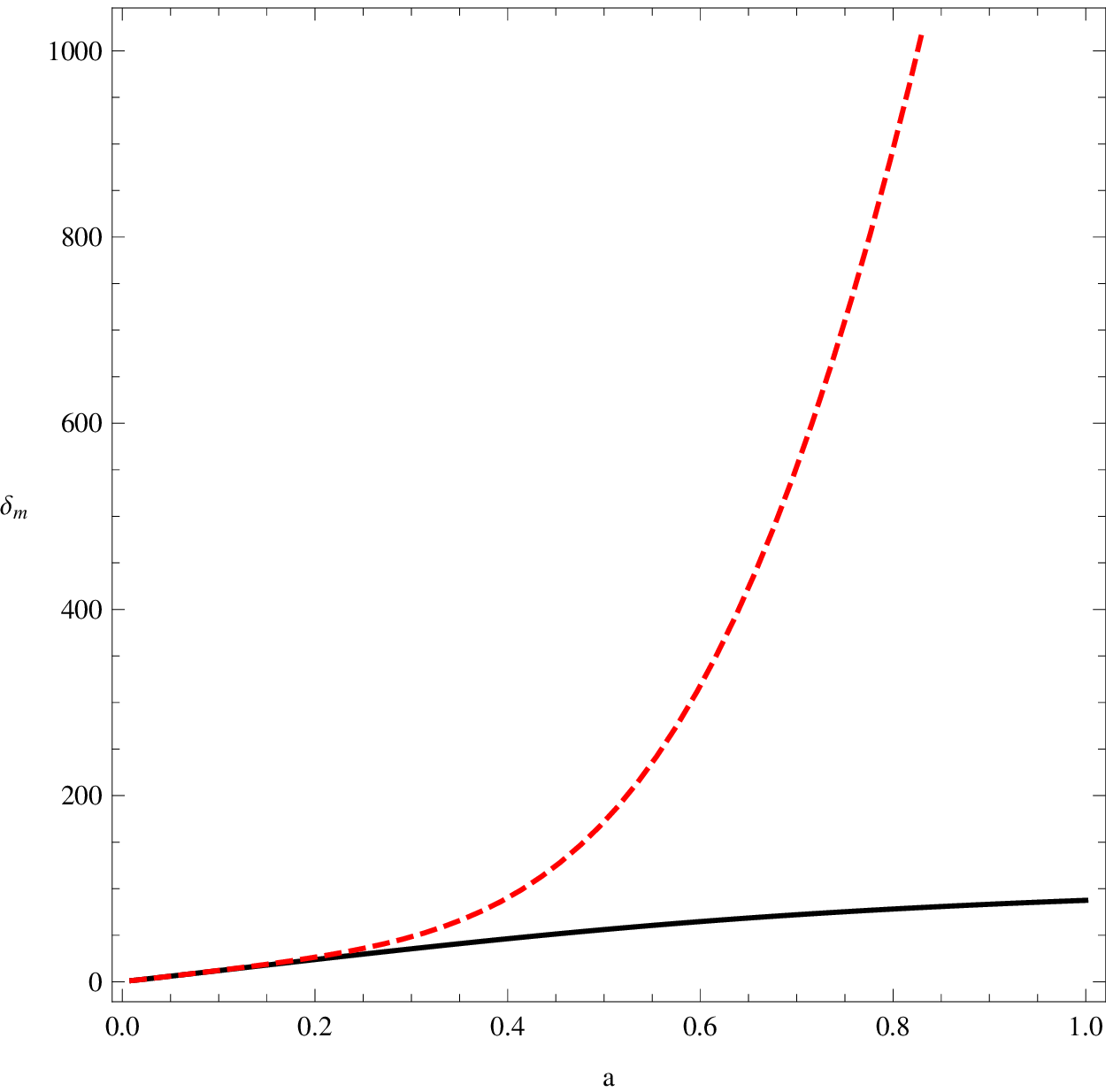}~\includegraphics[width=0.4\columnwidth]{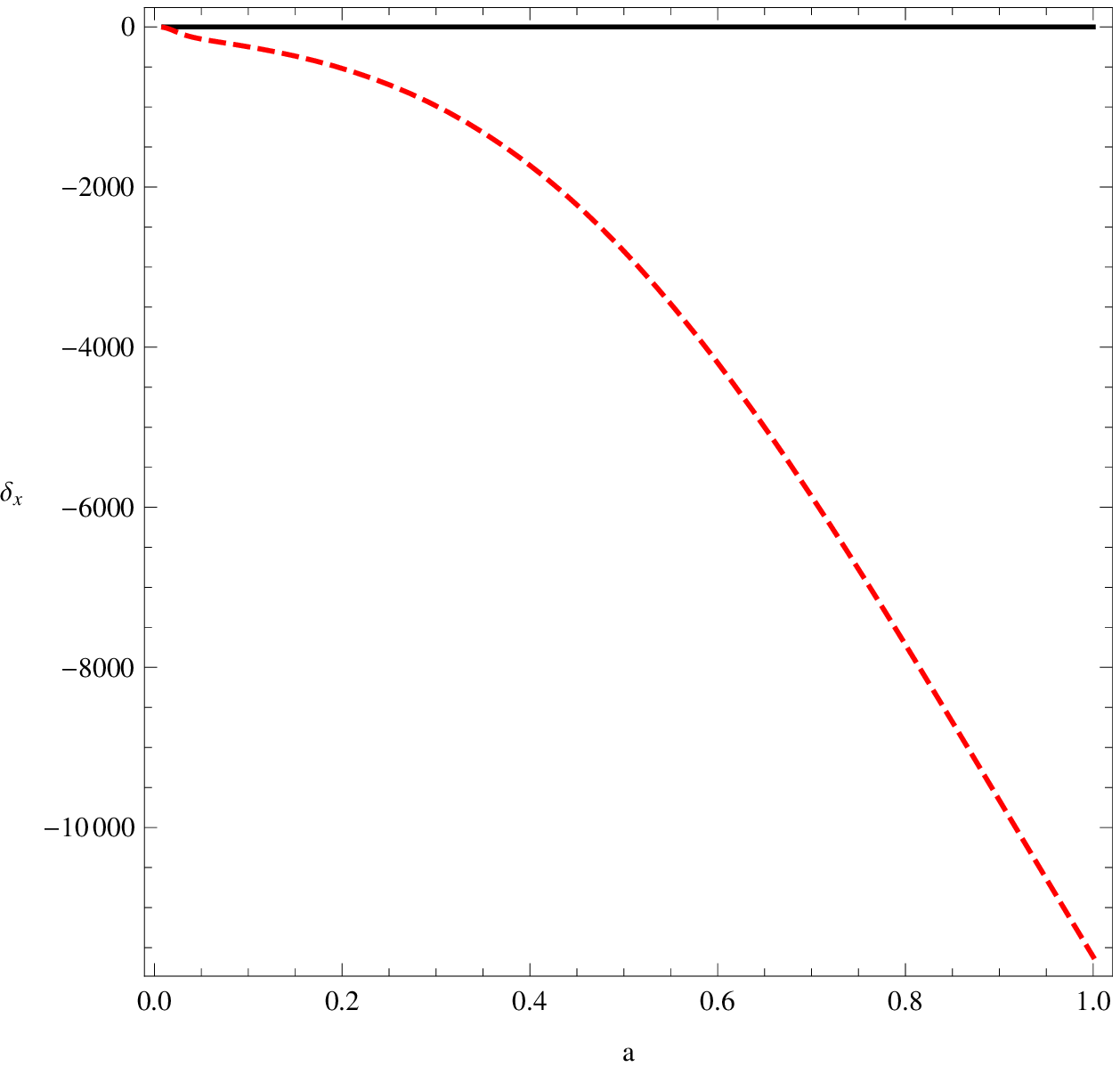}
\end{center}
\caption{Evolution of $\delta_m$ and $\delta_x$ as functions of the scale factor for the choice $\gamma = 1$ and $w_x = -1$ (black solid lines) and $\gamma - 1 = 10^{-4}$ and $1 + w_x = 10^{-4}$ (red-dashed lines).}
\label{fig2}
\end{figure}

For the equations governing the evolution of small perturbation, we follow \cite{Batista:2011nu}. We write the perturbed metric as $g_{\mu\nu}= g_{\mu\nu}^{(0)} + h_{\mu\nu}$, where $g_{\mu\nu}^{(0)}$ indicates the background flat FLRW metric of Eq.~\eqref{FLRWmet} and $h_{\mu\nu}$ is a small fluctuation. We choose the synchronous gauge condition, i.e. $h_{\mu0} = 0$ and introduce the perturbations as follows:
\begin{eqnarray}
\rho_{x} = \rho_{x}^{(0)}+\delta\rho_{x}\;, \quad \rho_{m}= \rho_{m}^{(0)} +\delta\rho_{m}\;,   \quad u_{m}=u_{m}^{(0)}+\delta u_{m}\;,
\label{r}\\
p_{x}= p_{x}^{(0)}+\delta p_{x}\;, \quad p_{m}= p_{m}^{(0)}+\delta p_{m}\;, \quad u_{x}=u_{x}^{(0)} + \delta u_{x}\;,\label{g}
\end{eqnarray}
In the expressions (\ref{r})-(\ref{g}), the superscript $(0)$ indicates the background functions and $\delta\rho_{x}$, $\delta\rho_{m}$, $\delta u_{m}$, $\delta u_{x}$, $\delta p_{x}$, $\delta p_{m}$, represent the perturbed quantities in density, four-velocity and pressure. We also introduce the following definitions:
\begin{equation}
\delta_{x} \equiv \frac{\delta \rho_{x}}{\rho_{x}}\;, \quad \delta_{m} \equiv \frac{\delta \rho_{m}}{\rho_{m}}\;, \quad \Theta \equiv \partial_{k}\delta u^{k}\;, \quad h \equiv \frac{h^{k}{}_k}{a^{2}}\;.
\end{equation}
For deeper detail, see for example \cite{Ma:1995ey}. After standard calculations, the perturbed conservation equation for the matter component can be cast in the following very simple form:
\begin{equation}
\dot{\delta}_{m}=\frac{\dot{h}}{2}\;.
\label{V}
\end{equation}
Rewriting Friedmann equation \eqref{1420} as
\begin{equation}
 \frac{H^2}{H_0^2} = \frac{\Omega_x}{2}[3-\gamma-3(1-\gamma)w_x] + \frac{\Omega_m}{2}(3-\gamma)\;,
\end{equation}
where
\begin{eqnarray}
\Omega_x &=& \Omega_{de0}a^{-\frac{6(1+w_x)}{2-(\gamma-1)(1-3w_{x})}} + \frac{3(1-\gamma)\Omega_{m0}}{6w_x + 3(\gamma-1)(1-3w_x)}a^{-3}\;,
\label{3b}\\
\Omega_{m} &=& \frac{\Omega_{m0}}{a^{3}}\;,
\end{eqnarray}
with $\Omega_{de0}$ and $\Omega_{m0}$ given in Eq.~\eqref{baa}, and calculating the perturbed component $R_{00}$, we get the following equation:
\begin{equation}
\ddot{\delta}_{m} + 2H\dot{\delta}_{m}-\frac{3H_0^2\gamma}{2}\Omega_{m}\delta_{m}=\frac{3H_0^2}{2}[\gamma+3(2-\gamma)w_x]\delta_{x}\Omega_{x}\;.
\label{4}
\end{equation}
Calculating the perturbation of (\ref{n}) we obtain for $\nu=0$
\begin{equation}
\delta\dot{\rho}_{x} + 3H(1+w_x)\delta\rho_{x}+ (1+w_x)\rho_{x}\biggr(\Theta_{x}-\frac{\dot{h}}{2}\biggl) =
\left( \frac{\gamma-1}{2}\right)[(1-3w_x)\delta\dot{\rho}_{x}+ \delta\dot{\rho}_{m}]\;.
\label{5}
\end{equation}
and for $\nu=i$
\begin{eqnarray}
(1+w_x)\dot{\rho}_{x}\Theta_{x}+ (1+w_x)\rho_{x}\dot{\Theta}_{x} + 5H(1+w_x)\rho_{x}\Theta_{x}= \nonumber \\ \frac{\nabla^{2}\delta\rho_{x}}{a^2}\left[\frac{1-\gamma}{2}+ \left( \frac{3\gamma-5}{2} \right)w_x\right]+ \frac{\nabla^{2}\delta\rho_{m}}{a^2}\left( \frac{1-\gamma}{2}\right)\;.
\label{8}
\end{eqnarray}
It is useful to rewrite the above equations in terms of the derivative of the scale factor $a$ instead of the cosmic time $t$ since the former is directly connected with the redshift via $z = -1 + 1/a$. In this way, Eq.~(\ref{4}) becomes
\begin{eqnarray}\label{deltameqa}
\delta^{\prime \prime}_{m}+ \delta^{\prime}_{m}\left(\frac{H'}{H} + \frac{3}{a}\right) - \frac{3H_0^2\gamma\Omega_{m}}{2H^2a^2}\delta_m &=&\frac{3H_0^2}{2H^2a^2}\left[\gamma + 3(2-\gamma)w_x\right]\delta_{x}\Omega_x\;,
\end{eqnarray}
Equation~(\ref{5}), using also Eqs.~\eqref{1419}, turns into
\begin{eqnarray}
\left[1 - \frac{\gamma - 1}{2}(1 - 3w_x)\right]\delta^{\prime}_{x} - \frac{3}{a}\frac{\Omega_m}{\Omega_x}\frac{\gamma - 1}{2}\delta_x + (1 + w_x)\frac{\Theta_x}{Ha}&=& \nonumber\\ \delta^{\prime}_{m}\left[\frac{\gamma - 1}{2}\frac{\Omega_{m}}{\Omega_{x}} + (1 + w_x)\right] - \frac{3}{a}\frac{\Omega_m}{\Omega_x}\frac{\gamma - 1}{2}\delta_m\;,
\label{13}
\end{eqnarray}
and finally Eq.~(\ref{8}) reads
\begin{equation}
\Theta^{\prime}_{x} + \left(\frac{\Omega^{\prime}_{x}}{\Omega_{x}} + \frac{5}{a}\right)\Theta_{x} = \frac{1}{1 + w_x}\frac{\nabla^{2}\delta_{x}}{Ha^3}\left[\frac{1-\gamma}{2} + \left(\frac{3\gamma-5}{2} \right)w_x \right] + \frac{1}{1+w_x}\frac{\Omega_m}{\Omega_x}\frac{\nabla^{2}\delta_{m}}{Ha^3}\left(\frac{1-\gamma}{2}\right)\;,
\label{10}
\end{equation}
where the prime denotes derivation with respect to the scale factor. In \figurename{~\ref{fig2}} we plot the evolution of $\delta_m$ and $\delta_x$ by solving the above equations from $z = 500$, with $\Omega_{m0} = 0.227$ and a scale $k = 0.001$ h Mpc$^{-1}$. As initial conditions we choose $\delta_m(z = 500) = 1$ and $\delta_x(z = 500) = 0$.

When $w_x = -1$, the evolution of $\delta_m$ and $\delta_x$ is independent of $\gamma$ and one reproduce exactly the $\Lambda$CDM model. This is the case also investigated in \cite{Batista:2011nu}. Note in the left panel of \figurename{~\ref{fig2}} that $\delta_x$ is identically vanishing, since it acts indeed as a cosmological constant, see Eq.~\eqref{3a}. A dramatic effect of instability appears as soon as $w_x$ differs from $-1$. In this case, if $\gamma$ slightly differs from unity, $\delta_m$ and $\delta_x$ rapidly grow of many order of magnitudes.

This may be ascribed to the existence of Eq.~\eqref{10} and the factor multiplying $\nabla^2\delta_{x}$, which may be regarded as an effective speed of sound, i.e.
\begin{equation}
 c_{sx}^2 \equiv \frac{\gamma - 1}{2} + \left(\frac{5 - 3\gamma}{2}\right)w_x\;.
\end{equation}
Indeed, when $\gamma = 1$, one recovers the GR result $c_{sx}^2  = w_x$. Therefore, we may speculate that the evolution of small perturbations in $\delta_x$ should not be affected by instabilities or oscillatory behaviors when $c_{sx}^2 = 0$, i.e. when
\begin{equation}
 \gamma = \frac{1 - 5w_x}{1 - 3w_x}\;,
\end{equation}
which, however, is the right limit of Eq.~\eqref{gammaconstr}, for which the denominator of the second term on the right hand side of Eq.~\eqref{3a} vanishes and therefore cannot be attained.


\section{Results and Discussion}\label{Sec:ResDisc}

In this work we investigate in some detail Rastall's theory in cosmology, on a flat Robertson-Walker metric, investigating a two-fluid model where one of the fluids is pressureless and obeys the usual conservation law (therefore scales as $a^{-3}$), whereas the other is described by an equation of state $p_x = w_x\rho_x$, with $w_x$ constant. This model is a generalization of the one studied in \cite{Batista:2011nu}, for which $w_x = -1$ and for which the equivalence with the $\Lambda$CDM at background and linear perturbations levels implies that no constraints on the parameter $\gamma$ can be established using the corresponding observational tests.
\par
We perform a Bayesian analysis on the type Ia supernovae Constitution dataset, and show that $\gamma$ is not strictly constrained whereas $w_x \sim -1$ is favored. On the other hand, considering the evolution of small perturbations we find a dramatic instability if $w_x \neq -1$ and $\gamma \neq 1$. This is a result which favors General Relativity, i.e. $\gamma = 1$. However, another possibility seems to be viable, i.e. $w_x = -1$, for which the dynamical equations turn independent of $\gamma$.
This is the case investigated in \cite{Batista:2011nu}, which thus seems to be the only possibility in which a hydrodynamical model could work in Rastall's cosmology. We expect different predictions involving $\gamma$ at the non-linear level of perturbations. A possibility to evade the strong constraints found in the perturbative analysis performed in the present paper is to describe
the dark energy component through a self-interacting scalar field. The non-linear analysis and the scalar field formulation of the Rastall's cosmological model are projects for future researches.

\section*{Acknowledgements}

We thank CNPq (Brazil) and FAPES (Brazil) for partial financial support. This research has made use of the CfA Supernova Archive, which is funded in part by the National Science Foundation through grant AST 0907903.



\begin{thebibliography}{99}
\bibitem{Li:2011sd}
  M.~Li, X.~-D.~Li, S.~Wang and Y.~Wang,
  ``Dark Energy,''
  Commun.\ Theor.\ Phys.\  {\bf 56} (2011) 525
  [arXiv:1103.5870 [astro-ph.CO]].
\bibitem{Caldwell:2009ix}
  R.~R.~Caldwell and M.~Kamionkowski,
  ``The Physics of Cosmic Acceleration,''
  Ann.\ Rev.\ Nucl.\ Part.\ Sci.\  {\bf 59} (2009) 397
  [arXiv:0903.0866 [astro-ph.CO]].
\bibitem{Bertone:2004pz}
  G.~Bertone, D.~Hooper and J.~Silk,
  ``Particle dark matter: Evidence, candidates and constraints,''
  Phys.\ Rept.\  {\bf 405} (2005) 279
  [hep-ph/0404175].
\bibitem{Clifton:2011jh}
  T.~Clifton, P.~G.~Ferreira, A.~Padilla and C.~Skordis,
  ``Modified Gravity and Cosmology,''
  Phys.\ Rept.\  {\bf 513} (2012) 1
  [arXiv:1106.2476 [astro-ph.CO]].
\bibitem{DeFelice:2010aj}
  A.~De Felice and S.~Tsujikawa,
  ``f(R) theories,''
  Living Rev.\ Rel.\  {\bf 13} (2010) 3
  [arXiv:1002.4928 [gr-qc]].
  \bibitem{Kamenshchik:2001cp}
  A.~Y.~.Kamenshchik, U.~Moschella and V.~Pasquier,
  ``An Alternative to quintessence,''
  Phys.\ Lett.\ B {\bf 511} (2001) 265
  [gr-qc/0103004].
\bibitem{Gorini:2007ta}
  V.~Gorini, A.~Y.~Kamenshchik, U.~Moschella, O.~F.~Piattella and A.~A.~Starobinsky,
  ``Gauge-invariant analysis of perturbations in Chaplygin gas unified models of dark matter and dark energy,''
  JCAP {\bf 0802} (2008) 016
  [arXiv:0711.4242 [astro-ph]].
 \bibitem{Piattella:2009da}
  O.~F.~Piattella,
  ``The extreme limit of the generalized Chaplygin gas,''
  JCAP {\bf 1003} (2010) 012
  [arXiv:0906.4430 [astro-ph.CO]].
  \bibitem{Piattella:2009kt}
  O.~F.~Piattella, D.~Bertacca, M.~Bruni and D.~Pietrobon,
  ``Unified Dark Matter models with fast transition,''
  JCAP {\bf 1001} (2010) 014
  [arXiv:0911.2664 [astro-ph.CO]].
  \bibitem{Bertacca:2010mt}
  D.~Bertacca, M.~Bruni, O.~F.~Piattella and D.~Pietrobon,
  ``Unified Dark Matter scalar field models with fast transition,''
  JCAP {\bf 1102} (2011) 018
  [arXiv:1011.6669 [astro-ph.CO]].
  \bibitem{Campos:2012ez}
  J.~P.~Campos, J.~C.~Fabris, R.~Perez, O.~F.~Piattella and H.~Velten,
  ``Does Chaplygin gas have salvation?,''
  arXiv:1212.4136 [astro-ph.CO].
  \bibitem{Zimdahl:1996ka}
  W.~Zimdahl,
  ``Bulk viscous cosmology,''
  Phys.\ Rev.\ D {\bf 53} (1996) 5483
  [astro-ph/9601189].
  \bibitem{Colistete:2007xi}
  R.~Colistete, J.~C.~Fabris, J.~Tossa and W.~Zimdahl,
  ``Bulk Viscous Cosmology,''
  Phys.\ Rev.\ D {\bf 76} (2007) 103516
  [arXiv:0706.4086 [astro-ph]].
  \bibitem{HipolitoRicaldi:2010mf}
  W.~S.~Hipolito-Ricaldi, H.~E.~S.~Velten and W.~Zimdahl,
  ``The Viscous Dark Fluid Universe,''
  Phys.\ Rev.\ D {\bf 82} (2010) 063507
  [arXiv:1007.0675 [astro-ph.CO]].
  \bibitem{Piattella:2011bs}
  O.~F.~Piattella, J.~C.~Fabris and W.~Zimdahl,
  ``Bulk viscous cosmology with causal transport theory,''
  JCAP {\bf 1105} (2011) 029
  [arXiv:1103.1328 [astro-ph.CO]].
  \bibitem{Zimdahl:2001ar}
  W.~Zimdahl and D.~Pavon,
  ``Interacting quintessence,''
  Phys.\ Lett.\ B {\bf 521} (2001) 133
  [astro-ph/0105479].
  \bibitem{Zimdahl:2005bk}
  W.~Zimdahl,
  ``Interacting dark energy and cosmological equations of state,''
  Int.\ J.\ Mod.\ Phys.\ D {\bf 14} (2005) 2319
  [gr-qc/0505056].
\bibitem{Rastall:1973nw}
  P.~Rastall,
  ``Generalization of the einstein theory,''
  Phys.\ Rev.\ D {\bf 6} (1972) 3357.
\bibitem{Rastall:1976uh}
  P.~Rastall,
  ``A Theory of Gravity,''
  Can.\ J.\ Phys.\  {\bf 54} (1976) 66.
\bibitem{Fabris:2011rm}
  J.~C.~Fabris, T.~C.~C.~Guio, M.~Hamani Daouda and O.~F.~Piattella,
  ``Scalar models for the generalized Chaplygin gas and the structure formation constraints,''
  Grav.\ Cosmol.\  {\bf 17} (2011) 259
  [arXiv:1011.0286 [astro-ph.CO]].
\bibitem{Batista:2011nu}
  C.~E.~M.~Batista, M.~H.~Daouda, J.~C.~Fabris, O.~F.~Piattella and D.~C.~Rodrigues,
  ``Rastall Cosmology and the $\Lambda$CDM Model,''
  Phys.\ Rev.\ D {\bf 85} (2012) 084008
  [arXiv:1112.4141 [astro-ph.CO]].
\bibitem{Fabris:2011wz}
  J.~C.~Fabris, M.~H.~Daouda and O.~F.~Piattella,
  ``Note on the Evolution of the Gravitational Potential in Rastall Scalar Field Theories,''
  Phys.\ Lett.\ B {\bf 711} (2012) 232
  [arXiv:1109.2096 [astro-ph.CO]].
  \bibitem{Daouda:2012ig}
  M.~H.~Daouda, J.~C.~Fabris and O.~F.~Piattella,
  ``Scalar models for the unification of the dark sector,''
  AIP Conf.\ Proc.\  {\bf 1471} (2012) 57
  [arXiv:1204.2298 [astro-ph.CO]].
\bibitem{Fabris:2012hw}
  J.~C.~Fabris, O.~F.~Piattella, D.~C.~Rodrigues, C.~E.~M.~Batista and M.~H.~Daouda,
  ``Rastall cosmology,''
  Int.\ J.\ Mod.\ Phys.\ Conf.\ Ser.\  {\bf 18} (2012) 67
  [arXiv:1205.1198 [astro-ph.CO]].
  \bibitem{Birrell:1982ix}
  N.~D.~Birrell and P.~C.~W.~Davies,
  ``Quantum Fields In Curved Space,''
  Cambridge, Uk: Univ. Pr. ( 1982) 340p
\bibitem{Fabris:1998hr}
  J.~C.~Fabris, R.~Kerner and J.~Tossa,
  ``Perturbative analysis of generalized Einstein's theories,''
  Int.\ J.\ Mod.\ Phys.\ D {\bf 9} (2000) 111
  [gr-qc/9806059].
\bibitem{Perivolaropoulos:2008ud}
  L.~Perivolaropoulos,
  ``Six Puzzles for LCDM Cosmology,''
  arXiv:0811.4684 [astro-ph].
\bibitem{Hicken:2009dk}
  M.~Hicken, W.~M.~Wood-Vasey, S.~Blondin, P.~Challis, S.~Jha, P.~L.~Kelly, A.~Rest and R.~P.~Kirshner,
  ``Improved Dark Energy Constraints from ~100 New CfA Supernova Type Ia Light Curves,''
  Astrophys.\ J.\  {\bf 700} (2009) 1097
  [arXiv:0901.4804 [astro-ph.CO]].
\bibitem{Riess:1998cb}
  A.~G.~Riess {\it et al.}  [Supernova Search Team Collaboration],
  ``Observational evidence from supernovae for an accelerating universe and a cosmological constant,''
  Astron.\ J.\  {\bf 116} (1998) 1009
  [astro-ph/9805201].
\bibitem{Capone:2009xm}
  M.~Capone, V.~F.~Cardone and M.~L.~Ruggiero,
  ``Accelerating cosmology in Rastall's theory,''
  Nuovo Cim.\ B {\bf 125} (2011) 1133
  [arXiv:0906.4139 [astro-ph.CO]].
  \bibitem{Ma:1995ey}
  C.~-P.~Ma and E.~Bertschinger,
  ``Cosmological perturbation theory in the synchronous and conformal Newtonian gauges,''
  Astrophys.\ J.\  {\bf 455} (1995) 7
  [astro-ph/9506072].
\end{thebibliography}
\end{document}